\documentclass[twocolumn,prl,amsmath,letterpaper,floatfix,showpacs]{revtex4}
\usepackage{graphicx}
\usepackage{amsmath}
\usepackage{amssymb}
\usepackage{color}

\newcommand{\Otwo}{{O$_{2}$ }}
\newcommand{\Ntwo}{{N$_{2}$ }}

\begin{document}

\title{Complete control, direct observation and study of molecular super rotors}

\author{Aleksey Korobenko$^{1}$, Alexander A. Milner$^{1}$ \& Valery Milner$^{1}$}

\affiliation{$^{1}$Department of  Physics \& Astronomy, The University of British Columbia, Vancouver, Canada}



\begin{abstract}
Extremely fast rotating molecules carrying significantly more energy in their rotation than in any other degree of freedom are known as ``super rotors''.
It has been speculated that super rotors may exhibit a number of unique properties. Theoretical studies showed that ultrafast molecular rotation may change the character of molecular scattering from solid surfaces\cite{Khodorkovsky11}, alter molecular trajectories in external fields\cite{Gershnabel10a}, make super rotors stable against collisions \cite{Tilford04}, and lead to the formation of gas vortices\cite{Steinitz12}. New ways of molecular cooling\cite{Forrey02} and selective chemical bond breaking\cite{Hasbani02} by ultrafast spinning have been suggested.
Bringing a large number of molecules to fast, directional and synchronous rotation is rather challenging\cite{Mullin95, Karczmarek99, Li00, Cryan11}. An efficient method of accelerating molecular rotation with an ``optical centrifuge'' has been proposed\cite{Karczmarek99} and successfully implemented\cite{Villeneuve00, Yuan11, Yuan11a}, yet only indirect evidence of super rotors has been reported to date.
Here we demonstrate the first direct observation of molecular super rotors and study their spectroscopic, dynamical and magneto-optical properties.
Using the centrifuge technique, we control the degree of rotational excitation and detect molecular rotation with high spectral and temporal resolution. Frequency-resolved detection enables us to map out the energy of extreme rotation levels, two orders of magnitude above the room temperature limit, and quantify the onset of the centrifugal distortion. Femtosecond time resolution reveals highly coherent rotational dynamics with lower de-coherence rates at higher values of the molecular angular momentum, and the increase of the molecular moment of inertia due to the rotation-induced chemical bond stretching. In the presence of an external magnetic field, ultrafast molecular rotation is found to result in an optical birefringence of the molecular ensemble.
We demonstrate that molecular super rotors can be created and observed in dense samples under normal conditions where the effects of ultrafast rotation on many-body interactions, inter-molecular collisions and chemical reactions can be readily explored.
\end{abstract}

\maketitle

Control of molecular rotation has been used for steering chemical reactions in gases\cite{Zare98} and at gas-surface interfaces\cite{Kuipers88, Shreenivas10}, for imaging individual molecular orbitals\cite{Corkum07} and generating extreme ultraviolet radiation\cite{Itatani05, Wagner07}, for deflecting molecular beams\cite{Purcell09} and separating molecular isotopes\cite{Fleischer06}. The appeal of rotational control has stimulated the development of multiple approaches in which molecules are exposed to strong non-resonant laser pulses\cite{Stapelfeldt03}. To extend the range of accessible angular momenta, a number of theoretical proposals aimed at guiding the molecules up the ``ladder'' of rotational levels ``step-by-step'' instead of exerting a single ultrashort rotational kick\cite{Karczmarek99, Li00, Vitanov04, Cryan11, Zhdanovich11}. Molecular spinning with an ``optical centrifuge'', suggested by Karczmarek \textit{et. al.} and implemented in this work, is achieved by forcing the molecules to follow the rotating polarization of a laser field\cite{Karczmarek99}. The final speed of rotation is determined by the spectral bandwidth of the laser pulse and may exceed $10^{13}$ revolutions per second. To make \Otwo molecules rotate primarily with this speed in thermal equilibrium, the gas temperature would have to be risen to above 50,000 Kelvin.

Since the original proposal\cite{Karczmarek99}, an optical centrifuge has been implemented by two experimental groups. In the pioneering work by Villeneuve \textit{et. al.}, dissociation of chlorine molecules exposed to the centrifuge field has been attributed to the breaking of the Cl-Cl bond which could not withstand the extremely high spinning rates\cite{Villeneuve00}. More recently, Yuan \textit{et. al.} observed rotational and translational heating in the ensembles of CO$_{2}$ and N$_{2}$O molecules and associated it with the collisional relaxation of the centrifuged species\cite{Yuan11, Yuan11a}. In both cases, an \textit{incoherent} secondary process (i.e. dissociation and multiple collisions) has been used for indirect identification of the formation of super rotors whose most unique property - their synchronous uni-directional rotation, remained hidden. In this work, we employ a \textit{coherent} detection technique, which enables us to observe the super rotors directly and to study the properties of these exotic molecular objects with high frequency and time resolution. In a state-resolved measurement of the rotational spectrum of oxygen, we reach the rotational quantum numbers as high as $N>100$ and analyze the effect of centrifugal distortion on the deviation of the rotational energy from that of a rigid rotor. In time-resolved detection, we follow the molecules as they spin up inside the rotating laser field and, after releasing them from the centrifuge, examine the influence of ultrafast rotation on the decay rate of their rotational coherence. Weak coupling between the electronic spin of \Otwo and the rotation of its nuclei, enhanced at high levels of rotational excitation, is investigated by placing the super rotors in an external magnetic field and observing the rotation-induced optical birefringence.
\begin{figure*}[tb]
\includegraphics[width=1.5\columnwidth]{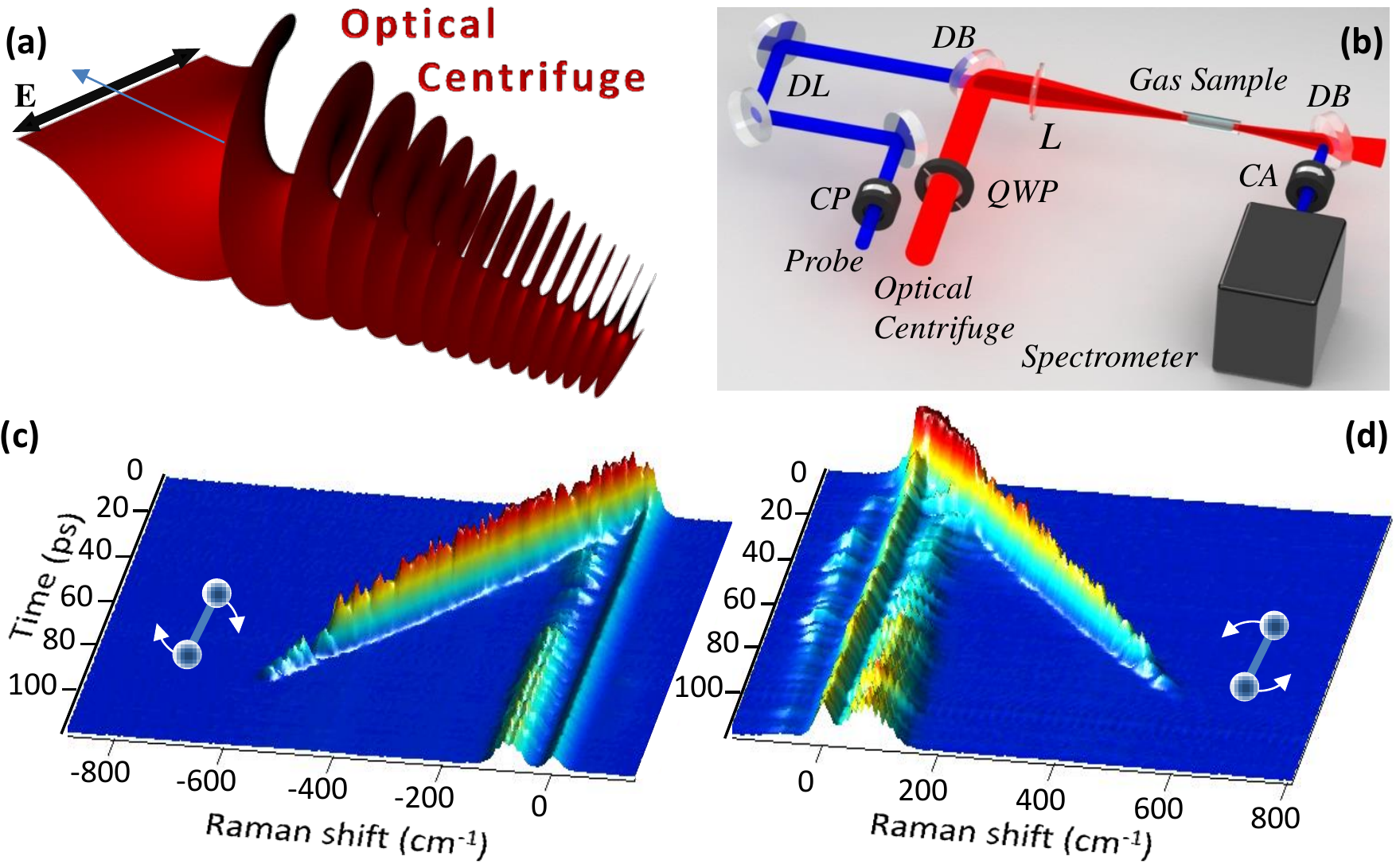}
\caption{\small{\textbf{a,} Illustration of an optical centrifuge field. The vector of linear polarization \textbf{E} undergoes an accelerated rotation, completing about 200 turns during a 70 ps centrifuge pulse and reaching frequencies on the order of $10^{13}$ Hz (for clarity, much lower angular acceleration is shown in the picture above). A laser-induced dipole force keeps the molecules aligned with \textbf{E}, accelerating them to the ultra high rotational velocities of a ``super rotor''. \textbf{b,} Experimental set up. Coherent Raman scattering is employed to determine the speed and direction of molecular rotation. Centrifuge pulses (red) are combined with a weak probe beam (blue) on a dichroic beamsplitter (DB) and focused into a gas cell by a 1 m-focal length lens (L) to a beam diameter of 120 $\mu $m (full width at half maximum, FWHM). Control of the spinning direction is executed by a quarter-wave plate (QWP). Probe pulses are polarized with a circular polarizer (CP), delayed with a delay line (DL) and scattered off the centrifuged molecules. After being filtered out with a second dichroic beamsplitter and a circular analyzer (CA), probe spectrum is recorded with a 0.1 nm-resolution spectrometer. Raman signals from the centrifuged Oxygen (color-coded according to the signal strength) are shown in panels \textbf{c} and \textbf{d} for the case of clockwise and counter-clockwise centrifuge spinning, respectively. As the molecules spend longer time in the centrifuge, the observed Raman frequency shift increases, providing a direct evidence of accelerated molecular rotation in one well defined direction.}}
\label{Fig-Setup}
\end{figure*}

Following the recipe of Karczmarek \textit{et. al.}\cite{Karczmarek99}, we create an optical centrifuge by generating a linearly polarized field with the polarization plane rotating along the propagation axis as illustrated in Fig.\ref{Fig-Setup}\textbf{a}. Accurate experimental characterization of the centrifuge field shows that the centrifuge angular frequency gradually increases from 0 to 10 THz in the course of about 70 ps. Adjusting the duration of the centrifuge pulse allows us to control the final speed of molecular rotation within these limits\cite{Korobenko13}. The centrifuge pulse is focused into a cell with molecular Oxygen or Nitrogen at room temperature and atmospheric pressure (Fig.\ref{Fig-Setup}\textbf{b}). Special care is taken to avoid ionization and plasma breakdown by limiting the peak intensity of the excitation field to below $5\times 10^{12}$ W/cm$^{2}$.

Key to this work is the use of coherent Raman scattering of probe light from the rotating molecules. Quantum mechanically, synchronous molecular rotation corresponds to a superposition of a few rotational quantum states - a ``rotational wave packet'', with an average frequency separation matching the frequency of the classical rotation. Owing to the time-dependent wave packet coherence, the probe spectrum acquires a frequency sideband shifted from the central probe frequency. The sign and the magnitude of the frequency shift $\Omega$, which can also be viewed as a result of the rotational Doppler effect\cite{Korech13}, indicates the direction and speed of the laser-induced molecular rotation, respectively\cite{Korobenko13}.
\begin{figure*}[t]
\includegraphics[width=1.75\columnwidth]{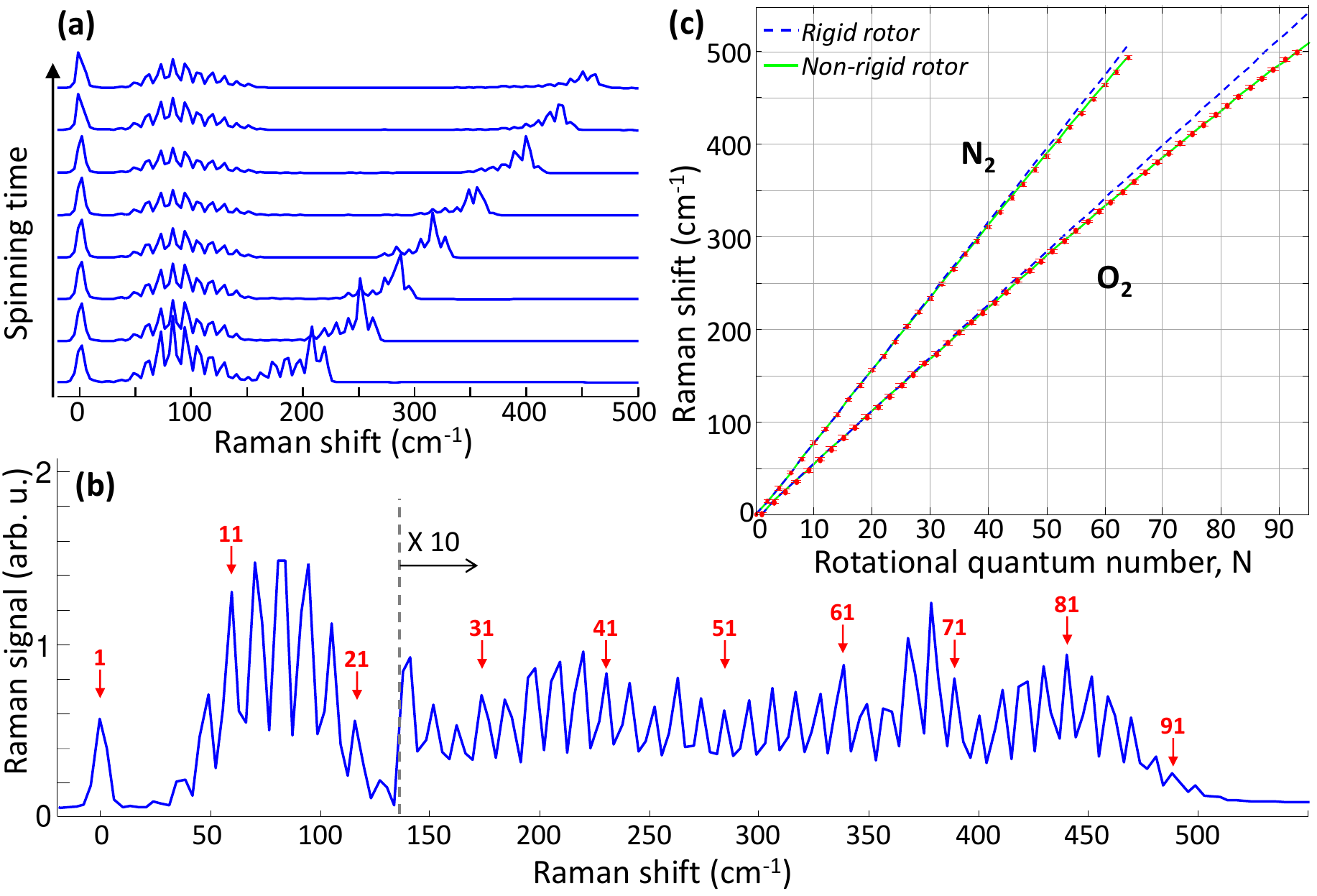}
\caption{\small{\textbf{a,} Raman spectra of the centrifuged Oxygen molecules, taken with narrowband (0.12 nm FWHM) probe pulses, as a function of the spinning time inside the centrifuge. The group of the Raman-shifted lines on the right side of the spectrum reflects an accelerated rotation of the molecules trapped in the centrifuge field, and demonstrates our ability to control the degree of rotational excitation. The sum of all measured spectra is shown in plot \textbf{b}. Well resolved peaks in the spectrum correspond to the individual Raman transitions between the states with rotational quantum numbers $N-2$ and $N$ (red labels, with ``1'' marking the unshifted probe spectrum). Only odd values of $N$ are allowed for $^{16}O_{2}$ molecule because of its nuclear spin statistics. \textbf{c,} Red dots with experimental error bars depict the rotational energy spectrum, expressed as a Raman frequency shift, for Oxygen and Nitrogen. Owing to the majority of the symmetric nuclear spin isomer of $^{14}N_{2}$ (ortho-Nitrogen) at normal conditions, we resolve only even $N$'s in its spectrum. Dashed blue lines are calculated in the ``rigid rotor'' approximation, whereas solid green curves are calculated with the centrifugal distortion taken into account (first two terms of the Dunham expansion\cite{NIST}). The calculations were carried out with no fitting parameters. The importance of the centrifugal distortion is evident at $N>50$.}}
\label{Fig-Frequency}
\end{figure*}

Delaying the arrival time of probe pulses with respect to the beginning of the centrifuge pulse enables us to observe the spinning molecules before and after they leave the centrifuge, following their dynamics for up to a few nanoseconds with sub-picosecond resolution. The ability to control the duration of probe pulses and their spectral bandwidth presents an opportunity for both time- and frequency-resolved detection. In Figures \ref{Fig-Setup} \textbf{c} and \textbf{d}, the rotational Raman signal from Oxygen is plotted as a function of time a molecule spent in the centrifuge. The growing frequency shift of the Raman sideband indicates the expected increase of the laser-induced spinning rate. The two plots correspond to two opposite senses of the centrifuge rotation and demonstrate the direct evidence of molecular spinning in both directions.

To identify the excited rotational levels, we narrow the probe bandwidth down to 7.5 cm$^{-1}$ (FWHM) and analyze the spectrum of the Raman signal. The results, corresponding to different centrifuge durations and, therefore, different degrees of rotational excitation, are shown in Figure \ref{Fig-Frequency} \textbf{a} for the case of \Otwo. Each measured spectrum consists of three parts: (i) the unshifted probe line at 0 cm$^{-1}$; (ii) the ``constant-speed wave packet'' centered around 80 cm$^{-1}$ and corresponding to the molecules kicked by the leading edge of the centrifuge but not trapped in it; and (iii) the ``accelerating wave packet'' corresponding to the centrifuged molecules. The observed fine structure stems from the discrete energy spectrum of a quantum rotor.
\begin{figure*}[tb]
\includegraphics[width=1.75\columnwidth]{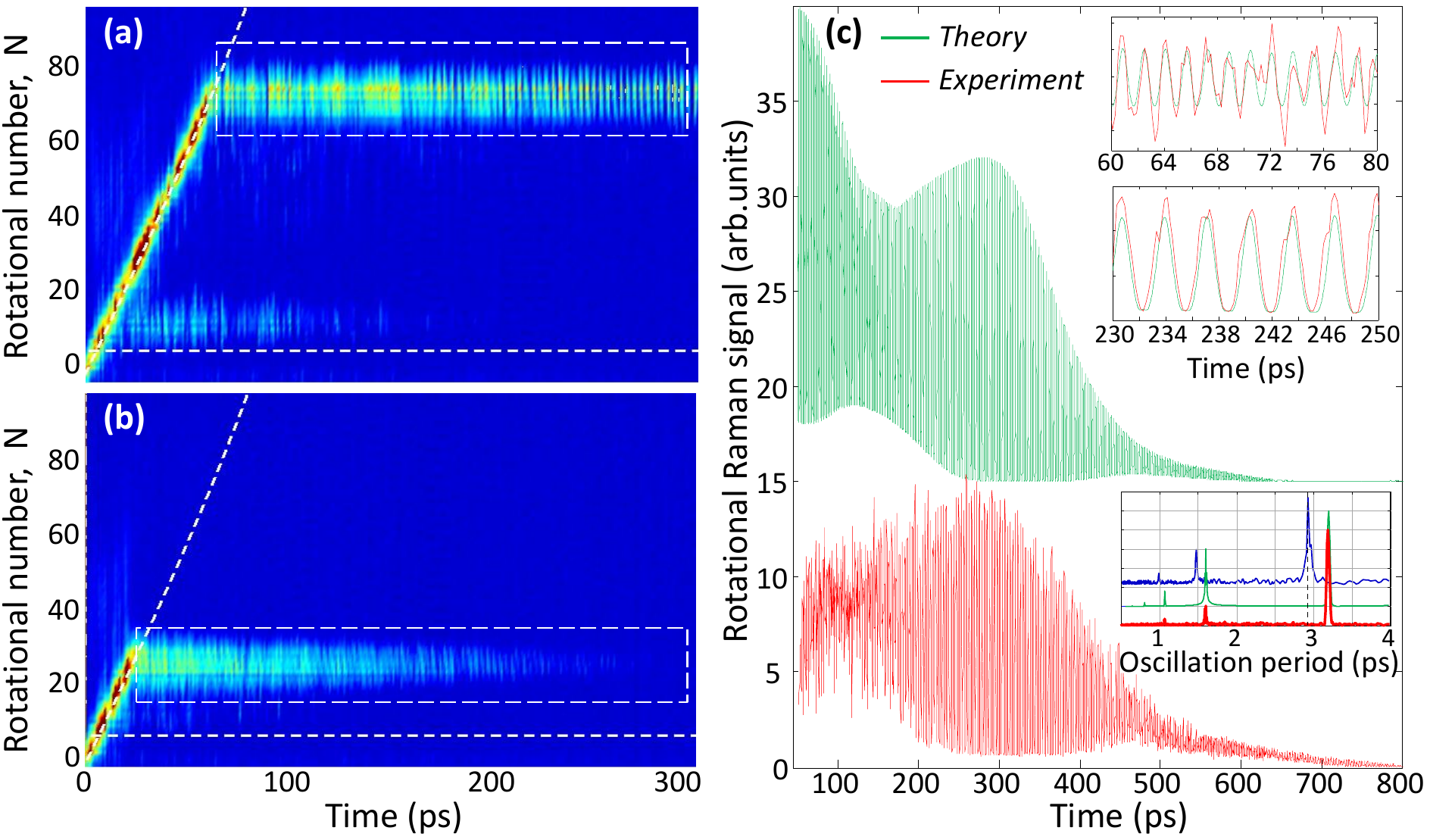}
\caption{\small{ \textbf{a,b} Selective centrifuge spinning of oxygen to $N\approx 69$ and $N\approx 29$, respectively. Until the time of release from the centrifuge, molecular rotation follows the angular frequency of the laser field, which is determined experimentally\cite{Korobenko13} and shown with dashed tilted lines. A horizontal Raman trace originating at the end of the centrifuge pulse (inside dashed rectangles at around 64 ps and 24 ps, respectively) indicates the free rotation of molecular super rotors. For comparison, dashed horizontal lines mark the most populated rotational state of \Otwo at room temperature, $N=7$. \textbf{c,} Experimentally detected (lower red) and numerically calculated (upper green) evolution of a super rotor wave packet centered at $N=69$ in Oxygen. Despite the lower experimental contrast at earlier times, associated with a marginal time resolution attained with 500 fs probe pulses, characteristic change of periodicity in the observed oscillations is well reproduced in the calculations (two upper insets), which were scaled to match the amplitude of the experimental signal. In agreement with the general theory of quantum wave packets\cite{Leichtle96,Seideman99}, the evolution of freely rotating molecules exhibits a discrete set of commensurate periods, revealed by a Fourier transform of the collected data (lower inset). In the case of a ``slow'' rotation ($N=29$, upper blue curve), the main period of 2.9 ps coincides with the prediction of the ``rigid rotor'' model (vertical dashed line). Centrifugal distortion of a fast super rotor ($N=69$) results in the stretching of the molecular bond and the correspondingly longer rotational period, as illustrated by the experimental results (bottom red line) and numerical calculations (middle green line).}}
\label{Fig-Time}
\end{figure*}

In a ``rigid rotor'' approximation, the energy of the state with the rotational quantum number $N$ is $E(N)=BN(N+1)$, where $B$ is the rotational constant of the molecule. This scaling results in a series of equidistant Raman peaks separated by $\Delta \Omega  = 4B\Delta N$, with $\Delta N$ being the smallest possible step in the molecular rotational ladder. Owing to the nuclear spin statistics, only odd values of $N$ are allowed in the spectrum of $^{16}$O$_{2}$, which dictates $\Delta N=2$ and therefore $\Delta \Omega \approx 11.4$ cm$^{-1}$. As shown in Figure \ref{Fig-Frequency} \textbf{b}, the detected peaks can indeed be resolved and assigned their respective rotational quantum numbers for up to $N=95$. The peak separation, however, does not stay constant, but rather decreases with increasing angular momentum - a direct consequence of the centrifugal distortion. The effect is demonstrated in Figure \ref{Fig-Frequency} \textbf{c} for both Oxygen and Nitrogen. Being able to resolve the energies of extreme rotational states, we quantify the magnitude of the centrifugal distortion and verify that it is well described by the Dunham expansion to second power in $N(N+1)$ [Ref.\cite{NIST}].

To study the rotational motion of the super rotors, we shorten the length of probe pulses to 500 fs and examine the time dependence of the Raman response. As shown in Figures \ref{Fig-Time} \textbf{a} and \textbf{b} for two levels of rotational excitation, molecules released from the centrifuge generate an oscillatory signal. The latter is indicative of the coherent rotation with well defined relative phase between the quantum states inside a rotational wave packet. Knowing the wave packet composition from the state-resolved detection discussed above, we model the dynamics numerically and compare it with the experimental results in Figure \ref{Fig-Time} \textbf{c}. As expected form the quadratic scaling of energy with the rotational quantum number $N$, the oscillations exhibit multiple commensurate time periods\cite{Leichtle96,Seideman99}. In the presented example for Oxygen at $N\approx 69$, the initial period of 1.6 ps changes to twice that value at later times, in excellent agreement with the numerical calculations (see two upper insets in Fig.\ref{Fig-Time}).
\begin{figure*}[tb]
\includegraphics[width=1.5\columnwidth]{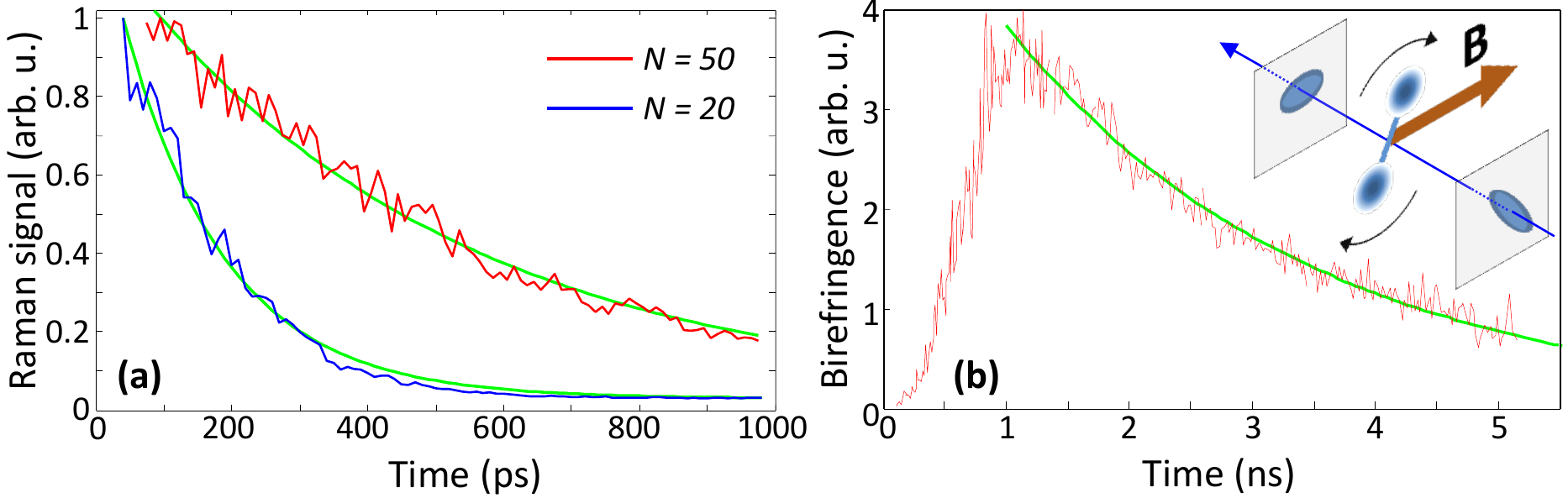}
\caption{\small{It is evident from plots \textbf{a} and \textbf{b} in Figure \ref{Fig-Time} that the decay of coherent molecular rotation is slower for molecules spinning with higher angular momentum. In panel \textbf{a}, this result is analyzed for Nitrogen super rotors spun to $N\approx 20$ (blue) and $N\approx 50$ (red). The observed falloff agrees well with an exponential decay. The de-phasing time for the fast rotating molecules is found to be more than 3 times longer than for their slower counterparts. \textbf{b,} Rotation-induced magneto-optical birefringence of Oxygen super rotors, analyzed by measuring the change of polarization of probe pulses passing through the ensemble of centrifuged molecules (red curve). Constant magnetic field of $2.8\pm 0.2$ T is applied perpendicularly to the axis of molecular rotation, as schematically shown in the inset (with the two ellipses representing the input and output polarization of probe pulses). An exponential fit (green line) indicates the decay time of about 2.5 ns.}}
\label{Fig-DecayBfield}
\end{figure*}

As follows from the general theory of quantum wave packets\cite{Leichtle96,Seideman99}, the oscillation periods are inversely proportional to the second derivative of $E(N)$ with respect to $N$. For a rigid-rotor model of Oxygen, this would result in the main period of $T=(8Bc)^{-1}\approx 2.9$ ps ($c$ being the speed of light in vacuum), marked by a vertical dashed line in the lower inset of Fig.\ref{Fig-Time} \textbf{c}. Centrifugal distortion pulls the atoms apart, lengthening the molecular bond and increasing the molecular moment of inertia \textbf{I}. Being inversely proportional to \textbf{I}, the rotational constant $B$ decreases, resulting in a longer period of oscillations at higher levels of rotational excitation, $T=\left[8Bc(1-6\epsilon N(N+1))\right]^{-1}$, where $\epsilon \equiv D/B\approx 3\times 10^{-6}$ is the ratio between the two rotational constants in the Dunham expansion\cite{NIST}. Confirmed by our time-resolved observations (lower inset in Fig.\ref{Fig-Time} \textbf{c}) and consistent with our frequency-resolved findings in Figure \ref{Fig-Frequency}, centrifugal distortion becomes important for understanding the spectroscopic and dynamical properties of Oxygen super rotors with $N>50$.

Comparing panels \textbf{a} and \textbf{b} of Figure \ref{Fig-Time}, we note that the amplitude of coherent Raman scattering decays slower for the faster rotating molecules. Using the frequency-resolved data for Nitrogen super rotors excited to ``low'' $N\approx 20$ and ``high'' $N\approx 50$, we plot the observed falloff of the corresponding rotational coherence in Figure \ref{Fig-DecayBfield} \textbf{a}. Fitting the results with an exponential decay (solid green lines), we arrive at the decoherence time of 150 ps for the slower molecules and 488 ps for the faster ones. Given the average time of 138 ps between the collisions of \Ntwo at room temperature and atmospheric pressure, this observation suggests that a single collision is sufficient to scramble the phase of a slow molecular rotor, while more collisions are needed to de-phase an ensemble of fast super rotors. Though full theoretical analysis of the collisional decay of super rotors is not yet available, we attribute our experimental results to the qualitative difference between the collisions of fast and slow rotating molecules. At $N=50$, the rotational period of \Ntwo is about 168 fs, more than three times shorter than the characteristic collision time of 560 fs, estimated as the ratio of the molecule's ``hard sphere diameter'' and its thermal velocity. At this limit, fast angular averaging may result in a smaller perturbation to the rotational phase in comparison with that of a slower rotator (here, $N=20$) for which the rotational period (419 fs) approaches the collision time scale.

Direct observation of controlled molecular rotation in extremely broad range of angular frequencies reported in this work paves the way for exploring novel science with molecular super rotors. In Figure \ref{Fig-DecayBfield} \textbf{b} we demonstrate a new type of magneto-optical birefringence induced by ultrafast molecular rotation. We place the centrifuged molecules in a constant magnetic field of $2.8\pm 0.2$ Tesla aligned perpendicularly to the rotation axis, and measure their optical birefringence. The latter is manifested by the change of polarization of probe pulses, illustrated by the two ellipses in the inset, passing through the sample. Although very pronounced in \Otwo, the effect has not been detected in Nitrogen, which points at the key role of spin-rotation coupling. At $N\approx 71$, used in this experiment, the interaction energy between the electronic spin of Oxygen and its nuclear rotation reaches 2.5 cm$^{-1}$ (Ref.\cite{HerzbergBook}), and may provide the mechanism for the observed phenomenon. The decay time of the rotation-induced birefringence (measured at 2.5 ns) is longer than the decoherence time discussed above. This is consistent with the qualitative picture in which the direction of the molecular rotation, rather than its phase, is responsible for the detected optical anisotropy. The demonstrated utility of our method offers the possibility of exciting new tests of kinematic, electrical, magnetic, acoustic, optical and reactive properties of molecular super rotors.

\begin{acknowledgements}
This work has been supported by the CFI, BCKDF and NSERC. We thank Gilad Hurvitz and Sergey Zhdanovich for their help with the experimental setup. We gratefully acknowledge stimulating discussions with John Hepburn, Ilya Averbukh, Yehiam Prior and Ed Grant.
\end{acknowledgements}


\end{document}